\documentstyle[12pt]{article}
\newcommand{\beqn}{\begin{equation}}
\newcommand{\eeqn}{\end{equation}}
\newcommand{\ga}{\mbox{$^{71}{Ga}$}}

\newcommand{\brn}{\mbox{$^{8}{B}~$}}
\newcommand{\be}{\mbox{$^{7}{Be}$~}}

\renewcommand{\ge}{\mbox{$^{71}{Ge}$}}

\begin{document}

\setcounter{page}{0}
\thispagestyle{empty}

\vspace*{-1in}
\begin{flushright}
CUPP-97/1\\
SINP/TNP/97-09\\
astro-ph/9709078
\end{flushright}
\vskip 20pt
\begin{center}
{\large\bf{EXCITED STATES OF $^{71}{Ge}$ ABOVE THE NEUTRON\\
EMISSION THRESHOLD AND SOLAR NEUTRINO CAPTURE RATES FOR Ga
DETECTORS}}\\
\vskip 20pt
{\it Kamales Kar \\
Saha Institute of Nuclear Physics, 1/AF, Bidhannagar,\\ Calcutta 700 064,
India.} \\
\vskip 5pt
{\it Sukhendusekhar Sarkar$^*$\\
Inter-University Consortium for DAE Facilities,\\ Calcutta Centre,
Sector - III, Block LB, Plot 8,\\ Bidhannagar, Calcutta 700 091,
India.}\\
\vskip 5pt
{\it Srubabati Goswami and Amitava Raychaudhuri \\
Department of Pure Physics, University of Calcutta,\\
92 Acharya Prafulla Chandra Road, Calcutta 700 009, India.}\\
\vskip 30pt 
{\bf ABSTRACT}
\vskip 20pt
\end{center}
In Gallium detectors for solar neutrinos, the capture rate due to
Gamow-Teller transitions to excited states of $^{71}Ge$ beyond the
neutron emission threshold is usually neglected. We make a model
calculation to estimate its effect and find that this yields an
additional contribution which may be as much as 0.4 SNU, even larger
than that from the Isobaric Analog State in $^{71}Ge$ reached by Fermi
transitions, which is normally included in the standard predictions.

\vskip 15pt

\parindent 0pt

$^*$Present address: Department of Physics, North Bengal University,
Raja Rammohanpur, West Bengal 734 430, India. 


\parindent 20pt

\newpage
Research in the area of the solar neutrino deficit problem has reached
a very exciting stage. On one side, several groups are working on
making both the theoretical as well as the observed neutrino capture
rates for the different detectors more precise. On the other side,
explanations of this shortfall invoking non-zero neutrino mass and
mixing in the two- and three-flavour scenarios are being pursued
vigourously. Out of the four detectors which have already published
data, measurements using Gallium (by the GALLEX and SAGE groups) have
the lowest threshold (0.233 MeV) and are able to capture neutrinos
produced through all the reactions in the sun, including the most
abundant $pp$ neutrinos.  The theoretical prediction for the total
neutrino capture rate in the Ga detectors is $132^{+7}_{-6}$ SNU (Solar
Neutrino Unit) using the Standard Solar Model (SSM) of Bahcall and
Ulrich \cite{bu} whereas it is $137^{+8}_{-7}$ in the SSM of Bahcall
and Pinsonneault \cite{bp95} which includes He and heavy element
diffusion. To calculate the total capture rates one integrates the
solar neutrino flux times the capture cross-section over the range of
neutrino energy and performs summation over all the neutrino sources.
Evaluation of the neutrino capture cross-section as a function of the
energy needs the knowledge of the Gamow-Teller (GT) strengths to
excited states of the daughter nucleus \ge. The theoretical
calculations of these GT strengths depend sensitively on the choice of
the interaction Hamiltonian \cite{oda}. The SSM predictions of ref.
\cite{bu} and \cite{bp95} use the GT transition strengths deduced from
the forward angle  $(p,n)$ reaction data \cite{krofcheck} from the
analogous hadronic  process $^{71}{Ga} {(p,n)} {^{71}{Ge}}$. The
neutron emission threshold in \ge~ is 7.4 MeV, and, in the prevalent
practice, one does not consider any $\nu $ capture which  excites
states above this energy through the GT operator. This is based on the
assumption that for such states the partial width for neutron emission
is much larger than the $\gamma$-decay width. In this letter we go
beyond this assumption and probe the effect that an inclusion of the
above states might have on the final solar neutrino capture rate. To
set this in perspective,  the Isobaric Analogue State (IAS) in \ge~ is
observed at an energy of 8.932 MeV and Champagne {\it et al.}
\cite{champen} measured the upper limit of the ratio of the gamma decay
width to the total width for the IAS to be about 10\%.  Based on this,
Bahcall and Ulrich \cite{bu} add 10\% of the capture rate  coming from
the Fermi strength and this turns out to yield 0.2 SNU for the solar
neutrinos.  But the $n$-emission from the IAS to low-lying states of
$^{70}Ge$ with $T=3$ is suppressed by isospin conservation. On the
other hand, for the states below the IAS ({\it i.e.}, below 8.932 MeV
of excitation in $^{71}Ge$) with $T = \frac{7}{2}$, connected by the GT
transition, there is no such inhibition due to isospin conservation and
hence the neutron emission probability is expected to be much larger.
In this work we consider the contribution, albeit small, coming from
GT-transitions to states above 7.4 MeV using information from
beta-delayed particle emission experiments. We make a model calculation
for this and to the best of our knowledge no earlier estimate of this
possible additional contribution has been made. Efforts are on for the
experimental measurement of the ratio of the gamma decay width to the
total width for these states.

The total capture rate $R$ for \ga~ is given by $R=\Sigma{R_{i}}$,
where $R_{i}$ is the capture rate for neutrinos from the ${i}^{th}$
source. $R_i$ is obtained by convoluting the calculated cross-section
with $\phi_{\nu_{i}}(q)$, the flux of neutrinos of energy $q$ from the
source $i$ \cite{jnb}. The normalisation of the latter is fixed by the
SSM under consideration and one has:
\begin{equation}
R_{i}~=~ C_{i} {\frac{\int \phi_{\nu_{i}}(q)~\sigma(q)~dq}
{\int\phi_{\nu_{i}}(q)~\sigma_{B}(q)~dq}}
\end{equation}
where  $\sigma$ and $\sigma_{B}$ denote the absorption
cross-sections as calculated from this work and as given in ref.
\cite{jnb} respectively. $C_{i}$ is the prediction in SNU for
neutrinos from the $i^{th}$ source according to the SSMs of ref.
\cite{bu} or \cite{bp95} depending on which model is used as the
reference point.

The capture rates $\sigma(E_{i})$ of neutrinos from the ground state of
\ga~ to a state of \ge~ at an excitation energy $E_{i}$ has two parts
-- one coming from the Fermi transition and the other from GT
transitions -- 
and can be written as,
\beqn
\sigma(E_{i}) = \sigma_{F}(E_{i}) + \sigma_{GT}(E_{i}) 
\eeqn
where $\sigma_{F}(E_{i})$ and $\sigma_{GT}(E_{i})$ are given by,
\beqn
\sigma_{F}(E_{i})~=~0.1 (\pi c^{3} {\hbar}^{4})^{-1}  
~{g_{V}}^{2} B_{F}(E_{i})p_{e}\epsilon F(Z,\epsilon)
\label{sigmaf}
\eeqn
\beqn
\sigma_{GT}(E_{i})~=~(\pi c^{3} {\hbar}^{4})^{-1} 
g_{A}^{2} B_{GT}(E_{i}) p_{e} 
\epsilon F(Z,\epsilon)
\label{sigmagt}
\eeqn
In eqs. (\ref{sigmaf}) and (\ref{sigmagt}) $g_{V}$,
$g_{A}$ are the vector and axial vector coupling constants, 
$p_{e}$ and $\epsilon$ are the momentum and energy of the emitted
electron  and $F(Z,\epsilon$) is the $\beta$-decay Fermi
function for which we use the analytic form given by Schenter
and Vogel \cite{schenter}.
$B_{F}(E_{i})$ and $B_{GT}(E_i)$ are the squared Fermi and GT matrix
elements for \ge~ states at energy $E_i$ respectively. 
The total Fermi strength of $(N-Z)$ is concentrated at $E_{IAS}$ ($
\sim \delta(E_{i} - E_{IAS})$) and the net contribution is proportional
to $B_{F}(IAS)$. For the GT strength distribution the usual
practice is to use $B_{GT}(E_{i})$s from $(p,n)$ reactions. Here we are
interested in the GT strength beyond the neutron emission threshold for
which no $(p,n)$ study results are available and we use a simple
theoretical model for the strength distribution based on the
configuration structures of the initial and final states which get
connected by the one body GT transition operator.  The model for the GT
strength distribution is based on a formalism constructed for
calculating the beta decay/electron capture rates in stellar conditions
in \cite{ffn} and used recently in \cite{ska}  for calculating the
neutrino capture rates for the \ga~ detectors. In this method, one uses
pure shell model configurations to describe the ground state of $Ga$ as
well as the ground and excited states of \ge.  For example the ground
state of \ga~ is taken as $(1f_{\frac{7}{2}})^8~(2p_{\frac{3}{2}})^3$
for the protons and a completely filled $fp$ shell for the neutrons.
Out of the eight possible GT transitions: (1) $2p_{\frac{1}{2}}(n)
\rightarrow 2p_{\frac{3}{2}}(p)$, (2) $2p_{\frac{1}{2}}(n) \rightarrow
2p_{\frac{1}{2}}(p)$, (3) $1f_{\frac{5}{2}}(n) \rightarrow
1f_{\frac{5}{2}}(p)$, (4) $1f_{\frac{5}{2}}(n) \rightarrow
1f_{\frac{7}{2}}(p)$, (5) $2p_{\frac{3}{2}}(n) \rightarrow
2p_{\frac{3}{2}}(p)$, (6) $2p_{\frac{3}{2}}(n) \rightarrow
2p_{\frac{1}{2}}(p)$, (7) $1f_{\frac{7}{2}}(n) \rightarrow
1f_{\frac{7}{2}}(p)$, (8) $1f_{\frac{7}{2}}(n) \rightarrow
1f_{\frac{5}{2}}(p)$ connecting the ground state of \ga~ to eight pure
configurations in
\ge, (4) and (7) are completely blocked. Each remaining configuration
has many individual states in it and the strength distribution to each
configuration gives rise to a resonance taken to be of the Gaussian
form in this model. The centroid energy of the strength Gaussians is
given by,
\begin{equation}
E_{cent} ~~ = ~~ \Delta E_{s.p.} ~~ + ~~ \Delta E_{2-body} ~~
              + ~~ \Delta E_{pairing}  
\end{equation}
where $\Delta E_{s.p.} $ and $\Delta E_{pairing} $ are the differences
in the total single particle energies and pairing energies,
respectively, of the excited and the ground state configurations of
$^{71}Ge$. $\Delta E_{2-body} $ is the contribution coming from the
2-body potential and is treated in a simplified TDA model as in
\cite{ffn}.  The total GT strength to each configuration is given by
the sum rule expression of the product of single particle transition
strength, the number of neutron particles and the fraction of proton
holes \cite{ska}. The widths ($\sigma$) of the strength Gaussians to
different configurations are all taken equal and is a free parameter of
the model. One finds that $\sigma$ = 1.55 MeV gives the total capture
rate as 132.83 SNU (137.17 SNU) close to the standard values according
to ref. \cite{bu} (\cite{bp95}). So, for the purpose of comparison, we
use the strength width of 1.55 MeV.

When one includes the contribution from all states beyond the neutron
threshold of 7.4 MeV in \ge~ the capture rate becomes
133.56 SNU (138.005 SNU), an increase of 0.73 SNU (0.84 SNU) over the
usual SSM prediction of ref. \cite{bu} (\cite{bp95}). 
But this is clearly an overestimation as these states mostly decay by
particle emission and not by gamma emission. To get a realistic
estimate we use information collected from beta delayed particle
emission experiments. With $\Gamma_{\gamma}, \Gamma_{\mu}$ being the
partial widths of gamma and particle decay respectively, the total
width $\Gamma = \Gamma_{\gamma} + \Gamma_{\mu}$. Denoting the particle
separation energy by $B_{\mu}$, ``$\Gamma_{\mu}/\Gamma$ is zero below
$B_{\mu}$ and remains small immediately above that, then increases in
the space of a few hundred keV to nearly its maximum value, rises more
slowly and finally decreases again at much higher excitations''
\cite{hardy}.  This indicates that $\Gamma_{\gamma}/\Gamma$ first falls
slowly, then very rapidly and then slowly goes towards zero. We model
this by a Gaussian fall-off and include that as a multiplicative factor
in the GT strength distribution beyond 7.4 MeV. The width of this
Gaussian, $\sigma _{\gamma}$, is varied in the range 0.3 -- 0.7 MeV,
consistent with beta-delayed neutron emission experiments. We exhibit
the Gamow-Teller strength distribution in Fig. 1 both without and with
the modulation beyond the particle emission threshold. The results of
\cite{champen} indicate that $\Gamma _{\gamma} / \Gamma $ for the IAS
falls off to 10\% at an excitation energy of 1.532 MeV with respect to
the neutron emission threshold. In the case under consideration the
suppression will be much larger due to lack of isospin constraints, as
noted earlier. For comparison, we choose $\sigma _{\gamma}$ = 0.5 MeV
which makes exp$(-x^2 /\sigma _{\gamma}^2)$ = 8.4 $\times 10 ^{-5}$ at
$x$ = 1.532 MeV. Modulating the contribution beyond $n$-emission in
this manner, the total capture rate is 133.09 (137.46) SNU, an increase
by 0.26 (0.30) SNU over the standard value using the SSM of ref.
\cite{bu} (\cite{bp95}). The contribution from the different sources
are given in Table 1, where we see that the increase comes almost
entirely from \brn~neutrinos.  We have checked that the increase in the
capture rate does not depend very sensitively on the parameter
$\sigma_{\gamma}$: varying $\sigma_{\gamma}$ from 0.3 to 0.7 MeV
changes the total neutrino capture rate from 132.99 (137.36) SNU  to
133.16 (137.55) SNU using the solar model of ref.
\cite{bu} (\cite{bp95}).

The work discussed so far assumes that the model space for the
distribution of the GT strength includes only the four $fp$ shell
orbitals, namely $1f_{7/2}$, $1f_{5/2}$, $2p_{3/2}$ and $2p_{1/2}$.
However, the experimental pick-up reaction data  suggest that the
$1g_{9/2}$ neutron orbital is about 10\% occupied \cite{mat} in the
ground state of \ga~ and one also sees that \ge~ has a low-lying
excited state of ${9/2}^+$ at an excitation of only 0.1984 MeV 
\cite{bhat}. Also, theoretical estimates using spectral distribution
methods and with different interaction Hamiltonians in the ($2p_{3/2}$,
$1f_{5/2}$, $2p_{1/2}$, $1g_{9/2}$) shell model space show that
the $1g_{9/2}$ orbit is about 16 - 20 \% occupied \cite{tress} in the
ground state of \ga. Inclusion of the $1g_{9/2}$ orbital in shell model
calculations makes the model space too large to handle. On the other
hand, using our schematic model of the GT strength distribution one can
extend the calculation to include it rather easily
\cite{ks}. Here the ground stae of \ga~ is a pure configuration with
the $fp$ neutron orbitals full except $1f_{5/2}$ which has 4 particles
instead of 6 and the two extra particles are put in $1g_{9/2}$. This
configuration gives rise to two additional transitions other than the
ones listed earlier, namely $1g_{9/2}(n) \rightarrow 1g_{9/2}(p)$ and
$1g_{9/2}(n) \rightarrow 1g_{7/2}(p)$. For the GT strength distribution
this will give rise to seven Gaussians and one half-Gaussian instead of
five Gaussians and a half-Gaussian considered earlier \cite{ks}. The
centroids of these extra strength Gaussians are evaluated in the same
manner and they are given the same width parameters as others. The
$\beta^{-}$ sum rule strength, $S_{\beta^{-}}$, increases in this space
to 221/7 from the earlier value of $3(N-Z)$ -- {\it i.e.}, 27 -- giving
rise to $S_{\beta^{+}}$ of 32/7. With this form of GT strength
distribution one can again investigate how much the neutrino capture
rate increases for the Ga detectors if one includes the contribution
beyond the neutron emission threshold in \ge.  Using the same model for
the ratio of gamma decay width to neutron emission width, one gets  the
capture rates in this $fpg$ model spaces as given in Table 2 for the
solar models of ref. \cite{bu} and \cite{bp95}. These numbers are for
the same value of the strength width parameter of 1.55 MeV and
$\sigma_{\gamma}$ = 0.5 MeV as in the $fp$ case. We see that the
increase in the total capture rate from contributions beyond the
$n$-emission threshold is 0.32 SNU for fluxes of ref. \cite{bu} and
0.36 SNU for the solar model of ref. \cite{bp95}.  Varying
$\sigma_{\gamma}$ from 0.3 MeV to 0.7 MeV changes the total rate from
133.182 (137.572) SNU to 133.386 (137.806) SNU using the solar model of
ref. \cite{bu} (\cite{bp95}).

A microscopic calculation of the ratio of the gamma decay width to the
total width is being attempted.  In conclusion, we stress that the
contribution to the neutrino capture rate of the excited states of \ge~
beyond the neutron emission threshold are not negligible and are
actually larger than the Fermi transition contributions usually
included. Taking into account the contribution beyond the $n$-emission
threshold in \ge~ along with the extension of the space to $fpg$ orbits
give an increase as large as 0.37 SNU over the conventional theoretical
calculation done including the $fp$ shell orbits only.

\parindent 0pt

{\bf ACKNOWLEDGEMENTS}: This work has been supported by the Eastern Centre
for Research in Astrophysics, India. The research of A.R. is also
supported in part by the Council of Scientific and Industrial Research,
India and the Department of Science and Technology, India while S.G.
acknowledges a fellowship from the former agency.
\newpage
\begin{description}
\item{Table 1:} Neutrino capture rate on \ga~ without (columns A and C)
and with (columns B and D) contributions above the neutron emission
threshold using Bahcall-Ulrich and Bahcall-Pinsonneault solar neutrino
fluxes. Here $\sigma_{\gamma}$ has been chosen to be 0.5 MeV.
\end{description}
\[
\begin{array}{|c|c|c|c|c|} \hline
{\rm Solar~ Neutrino~} & \multicolumn{2}{c|}{\rm Bahcall-Ulrich~
SSM~\cite{bu}} & \multicolumn{2}{c|}{\rm
Bahcall-Pinsonneault~SSM~\cite{bp95}} \\ \cline{2-3} \cline{4-5} 
{\rm source} & {\rm~~~~~~~A~~~~~~~} & {\rm B} & {\rm ~~~~~~~C~~~~~~~}
& {\rm D} \\ \hline
{pp} & 67.781 & 67.781 & 66.727 & 66.727 \\ 
{pep} & 5.211 & 5.211 & 5.211 & 5.211 \\
{hep} & 0.043 & 0.045 & 0.043 & 0.045 \\
{\be} & 34.109 & 34.109 & 37.490 & 37.490 \\
{\brn} & 11.335 & 11.593 & 13.036 & 13.333 \\
{^{13}{N}} & 4.823 & 4.823 & 4.823 & 4.823 \\
{^{15}{O}} & 9.433 & 9.433 & 9.742 & 9.742  \\
{^{17}{F}} & 0.093 & 0.093 & 0.093 & 0.093 \\ \hline
{\rm Total} & {132.828} & {133.088} & {137.165} & {137.464} \\ \hline
\end{array}
\]
\newpage

\begin{description}
\item{Table 2:} Same as in Table 1 but for the $fpg$ case.
\end{description}
\[
\begin{array}{|c|c|c|c|c|} \hline
{\rm Solar~ Neutrino~} & \multicolumn{2}{c|}{\rm Bahcall-Ulrich~
SSM~\cite{bu}} & \multicolumn{2}{c|}{\rm
Bahcall-Pinsonneault~SSM~\cite{bp95}} \\ \cline{2-3} \cline{4-5} 
{\rm source} & {\rm~~~~~~~A~~~~~~~} & {\rm B} & {\rm ~~~~~~~C~~~~~~~}
& {\rm D} \\ \hline
{pp} & 67.781 & 67.781 & 66.727 & 66.727 \\ 
{pep} & 5.211 & 5.211 & 5.211 & 5.211 \\
{hep} & 0.044 & 0.047 & 0.044 & 0.047 \\
{\be} & 34.109 & 34.109 & 37.490 & 37.490 \\
{\brn} & 11.485 & 11.798 & 13.208 & 13.568 \\
{^{13}{N}} & 4.823 & 4.823 & 4.823 & 4.823 \\
{^{15}{O}} & 9.433 & 9.433 & 9.742 & 9.742  \\
{^{17}{F}} & 0.093 & 0.093 & 0.093 & 0.093 \\ \hline
{\rm Total} & {132.979} & {133.295} & {137.338} & {137.701} \\ \hline
\end{array}
\]

\newpage
\begin{center}
FIGURE CAPTION
\end{center}
Fig.1. The Gamow-Teller strength, $B_{GT}$, (solid line) as a function
of the excitation energy of the $Ge$ nucleus. Also shown is the
neutron-emission threshold in $Ge$ at 7.4 MeV. Modulating the
distribution beyond the particle emission threshold by a Gaussian
suppression factor of width 0.7 MeV (0.3 MeV) results in the
long-dashed (short-dashed) curve.

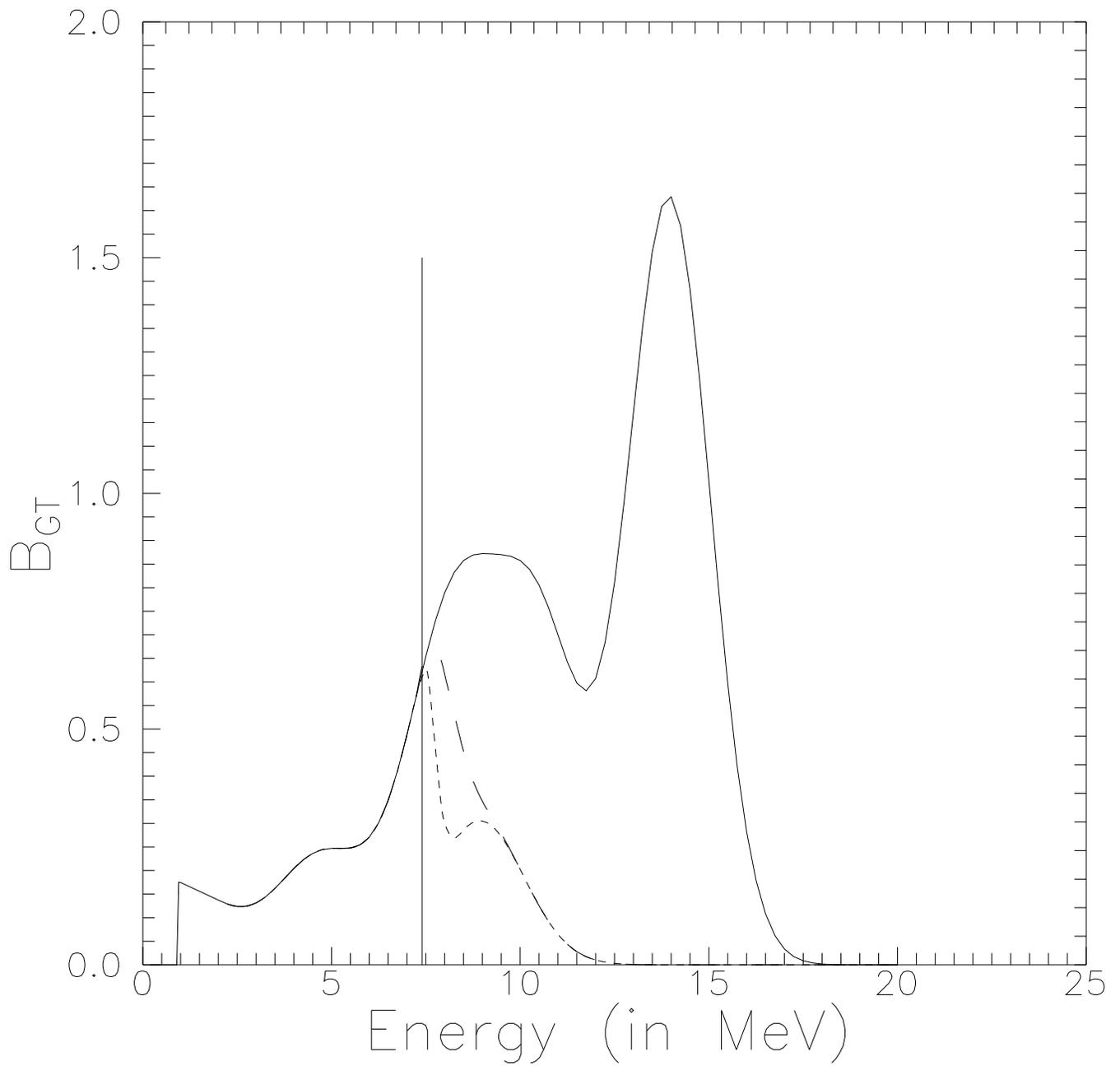
\end{document}